\documentclass[11pt,A4]{article}
\usepackage[english]{babel}
\usepackage[nottoc]{tocbibind}
\usepackage[utf8]{inputenc}
\usepackage{array}
\usepackage{float}
\usepackage{graphicx}
\usepackage{hyperref}
\usepackage{soul, color}
\sethlcolor{green}

\usepackage{listings}

\definecolor{light-gray}{gray}{0.97}

\usepackage[margin=1.1in]{geometry} 

\graphicspath{ {./figures/} }

\title{Toward equipping Artificial Moral Agents with multiple ethical theories}
\author{George Rautenbach and C. Maria Keet}
\date{Computer Science Department, University of Cape Town, South Africa\\
{\tt george@rauten.co.za}, {\tt mkeet@cs.uct.ac.za}}

\begin{document}
\maketitle

\begin{abstract}
Artificial Moral Agents (AMA's) is a field in computer science with the purpose of creating autonomous machines that can make moral decisions akin to how humans do. Researchers have proposed theoretical means of creating such machines, while philosophers have made arguments as to how these machines ought to behave, or whether they should even exist.

Of the currently theorised AMA's, all research and design has been done with either none or at most one specified normative ethical theory as basis. This is problematic because it narrows down the AMA's functional ability and versatility which in turn causes moral outcomes that a limited number of people agree with (thereby undermining an AMA's ability to be moral in a human sense). As solution we design a three-layer model for general normative ethical theories that can be used to serialise the ethical views of people and businesses for an AMA to use during reasoning. Four specific ethical norms (Kantianism, divine command theory, utilitarianism, and egoism) were modelled and evaluated as proof of concept for normative modelling. Furthermore, all models were serialised to XML/XSD as proof of support for computerisation. 
\end{abstract}

%\tableofcontents
%
%\tableoffigures

\section{Introduction}
An AMA is a computerised entity that can act as an agent and make moral decisions \cite{Allen_2000}. %(Allen et al., 2000). 
Computerisation simply entails that the agent is entirely comprised of software programs and hardware circuitry. Moral agency, on the other hand, is the philosophical notion of an entity that both has the ability to make a moral decision freely and has a kind of understanding of what each available action means in its context. 

Multiple proposals have been put forward for creating this artificial agency, notably \cite{anderson2005towards,Anderson_2018,benzmuller2018deontic,Liao_2019,sunstein1986legal}, which, however, take a single-theory approach.
That is, as basis for their AMA's reasoning design, the authors either pick one ethical theory from moral philosophy, or they choose some other goal that is morality-adjacent, but not necessarily moral (e.g., maximising agreement of stakeholders). 
Moral philosophers also discuss different ethical theories and which is best to use for AMA's (e.g., \cite{Allen_2000,lichocki2011ethical}).
None of the aforementioned works considers the possibility of using more than one theory for the same AMA. 

AMA's designed with one specific ethical theory as foundation has the advantage of deterministic behaviour but comes with two major problems. The first is the rigidity of having a single ethical theory as basis. By reasoning in only one specific way for every situation, the AMA is inadvertently opposing people who hold different views. 
For instance, whether an AMA-equipped chocobot should bake and serve yet more chocolate sweets to an already obese owner who is requesting another batch of rum-and-raisin chocolate muffins in an attempt to eat away life's sorrows: an AMA programmed with ethical egoism would do it because it is good for the owner's physiological satisfaction (albeit short-term), whereas a utilitarian chocobot would not since fattening already obese people with even more unhealthy food is not in society's best interest due to the increased healthcare costs. A utilitarian chocobot would make the obese owner unhappy, whereas an egoist chocobot would make a caring society unhappy.
Having a system fundamentally compatible with only one ethical norm is naturally bound to cause controversy and divide across people \cite{lichocki2011ethical}. %(Łichocki et al., 2011). 
In order to give people (specifically those affected by AMAs’ moral actions) an appreciation for computer agency, they must have their views considered too. After all, if the purpose of morality is human flourishing (which most moral philosophers agree it is)\footnote{We do not consider non-human or objective morality in this paper.}, then an AMA ought to take humans and their views into account. This means being able to consider one of multiple normative ethical theories during moral evaluation time. 

The second common problem is the infeasibility of storing sufficient information about relevant people and entities. By definition, to be a moral agent, the subject must have an understanding of the actions available to it. This invariably means the AMA should take into consideration the effects its actions will have on people. To do this properly, the executing AMA must have sufficient moral information on stakeholders to be able to consider them fully. What constitutes as “sufficient” is debatable, but at the very least, the AMA must know what the stakeholders’ moral views are so that it can calculate how they may be affected (by how much the action is in line with their views). To be able to reason under any circumstances, an AMA would have to store all moral values of all potentially relevant stakeholders. Depending on one’s definition of moral values, this could mean millions of datapoints per stakeholder. A moral value can be as broad as, e.g., “do onto others what you want done onto you” or as specific as e.g., “abortion is permissible only in the first 8 weeks of pregnancy”. Nuance in people’s views vary immensely and so a naive “sufficient” system of moral values will hold gigabytes of moral data per stakeholder
to consider in reasoning toward an answer, 
which is  
computationally and economically very expensive.

In this  
technical report
we address these problems by creating a general multi-layered model for ethical theories. This allows for a standardised way to define any ethical theory in a manner that an AMA can process and use in reasoning. Such a model resolves the rigidity problem by providing easy switching between alternative ethical theories for a correctly-designed AMA, as well as the computational expensive problem by succinctly modelling theories such that they are general enough to take up only a few kilobytes of space, but specific enough to model a person or business’s true ethical theory. To ensure the model is also computer-readable, it is serialised to XSD/XML. 

Next, we shall discuss some necessary background information in the field of normative ethics (Section~\ref{sec:prelim}). Afterwards, examination of related works will be done to show that the problems mentioned above apply to all of them (Section~\ref{sec:relwork}). The core of the paper follows, which is the detailing of the general ethical theory modelling process (Section~\ref{sec:theo}). The last part of the paper will cover evaluation of the model by virtue of some use cases (Section~\ref{sec:eval}), a discussion of the value and consequence of our work (Section~\ref{sec:disc}), and finally a conclusion to the paper (Section~\ref{sec:concl}). 

\section{Background}
\label{sec:prelim}

This section contains an overview of normative ethics in moral philosophy for readers that are not well-versed in the topic. The reader may skip this section if the background knowledge is known.

A normative ethical theory is – simply put – a way to live. It is a sort of guide for action to achieve fulfilment as a human. It defines what kinds of actions or states of being is right and wrong. There are many different normative ethical theories in the world that are lived out by people, and every one of them has their own justification. 

The way normative theories categorise right and wrong acts can be used to divide them into two kinds: consequentialist and non-consequentialist. Consequentialist theories (sometimes called teleological) use the consequences of actions to derive responsibilities (the ends justify the means). Non-consequentialist theories (sometimes called deontological) use some other method to argue that actions themselves have moral value (the means justify the ends). E.g., for a deontologist it may be wrong to kill a man because killing itself is wrong, while for a consequentialist it may be justified if that man was on his way to kill ten others. 

\subsection{Four common moral theories summarised}

We describe  some popular ethical theories that are frequently mentioned in moral philosophy papers and are important to comprehend to understand the related works of AMA's as well as for the purposes of our model designs. 

\subsubsection*{Utilitarianism}
    
    An action is right if out of its alternatives it is the one that increases overall happiness at the smallest emotional cost. The theory as originally proposed by Jeremy Bentham places human happiness at the centre of morality. There are however many critiques based on the vagueness of “happiness”, to which utilitarians respond by redefining it as preference-satisfaction. That is, making people's desires reality. This theory essentially makes it an agent’s imperative to increase happiness for all. 
    
\subsubsection*{Egoism}
    
    An action is right if its ends are in the agent's self-interest. That is, it increases the agent’s preference-satisfaction. This functions the same as utilitarianism but considers exclusively the agent’s happiness and no other people. This may seem intuitively immoral, but proponents argue that if one takes an objective approach to what is in one's self-interest, one will not harm others (because of societal punishment) and will do good to humanity (for e.g., societal praise).
    
\subsubsection*{Hedonism}
    
    An action is right if it increases pleasure and decreases pain. This theory takes a similar approach to utilitarianism, but does not attempt to assign morality to any kind of intuitive true fulfilment. Rather it proposes that an agent focuses all their energy on maximising somatic pleasure and chasing evolutionary dopamine rewards (e.g., eating, mating, and fighting). This is typically the sort of life most animals lead, but hedonists argue it can be fit for humans too. 
    
\subsubsection*{Divine command theory}
    
    An action is right if God wills it so. Divine command theorists typically hold that God plays a three-fold role in ethics: Epistemic (God provides knowledge of morality), ontological (God gives meaning to morality, or morality cannot be defined otherwise), and prudential (God motivates humans to be moral by offering eternal rewards to those that are and eternal punishment to those that are not). In this paper we will use the Christian version of DCT, which is based on The Bible’s ten commandments. There are some variations in Christianity, which may affect what or who is relevant for moral consideration; we take an inclusive approach henceforth.
    
\subsubsection*{Kantianism}
    
    An action is right if it is universally willable and it respects the rational autonomy of others. Immanuel Kant proposes a very complex theory of morality that involves two main formulations of what he calls categorical imperatives (universal duties): the formulation of universal law, and the formulation of humanity. The former says that one should imagine what a world would be like where an action is law, and then extrapolate from such a state to determine the action’s morality (if the world is unlivable, the action is wrong). The latter says that one should always treat other people as ends in themselves, never as mere means to an end. Both formulations strive to cultivate respect for other people’s rational autonomy (that is, never to inhibit another’s ability to make a free decision). 
    
    Many criticise Kantianism and other deontological theories for being consequentialism in disguise (how do you know if a world is liveable but to check its consequences), but the key difference is that where deontology uses consequences it uses them to judge actions based on universal effects. That is, build a theoretical world where a certain condition applies to all agents and evaluate what that world would be like. This is why deontological theories implemented the way Kant would want it are very computationally expensive\footnote{For a more computer-friendly implementation of deontology, see \cite{benzmuller2018deontic}.}. There are also deontological theories that judge actions based on simple rules that have no human justification, e.g., Christian DCT employs the ten commandments as moral principles with justification limited to it being God’s word. 
    On the other hand, consequentialism focuses on the real-world outcomes of the specific action to be evaluated. In summary, as far as consequences are concerned, consequentialism leverages situational outcomes and deontology leverages universalised outcomes. Sometimes an action’s morality will be judged the same by both kinds of theories, but the justifications will always be different.\\

There are many different ethical theories, but in this paper we only consider the two most popular of each type to show that the model is sufficiently well-defined\footnote{Note that all these theories have well argued-for justifications to them that for sake of conciseness we do not explore in this paper. It is precisely because of these good rationales that it is worth considering multiple theories for AMA’s, and not just one.}: For consequentialist theories we use utilitarianism and egoism, and for deontological theories we use Kantianism and divine command theory. 

Table~\ref{table:1} shows a concise summary of the four relevant ethical theories. 

\begin{center}
\begin{table}[h!]
\caption{Summary of the normative ethical theories modelled in this paper.}
\centering
\begin{tabular}{p{2.8cm} || p{2.5cm} | p{2.1cm} | p{2.1cm} | p{2.1cm}} 
 \hline
  & Utilitarianism & Egoism & DCT & Kantianism \\
 \hline\hline
 Consequentiality & True & True & False & False \\ 
 \hline
 Patients & Humans & Humans & All life & Humans \\
 \hline
 Principles & Preference satisfaction for all & Preference satisfaction for the agent &
 Ten commandments & Categorical imperative
\end{tabular}
\label{table:1}
\end{table}
\end{center}

\subsection{Some common moral dilemma cases}
\label{sec:cases}

While there are several list of interesting cases for moral dilemmas, we highlight three that are relevant for the digital assistants or AMAs and thus which any software application ideally should be able to handle.

\subsubsection*{Case 1: The trolley problem}

In an applied ethics paper on abortion,  Philippa Foot introduced an ethical dilemma thought experiment known as the Trolley Problem \cite{foot1967problem}. It has since gained a lot of traction and is notoriously controversial in moral philosophy. This problem is included as a use case  
with the purpose of showing usefulness in well-known moral dilemma scenarios that are not AMA-relevant by nature, but that an AMA can find itself in nonetheless. An adapted version of the trolley problem follows. 

\begin{quote}
A train’s braking system fails, and it hurls at full speed towards an unwitting group of five people standing on the tracks. The automatic train track management system detects this and has the opportunity to switch tracks ahead of the train’s collision with the group. However, doing so will direct the wild train onto a track which is currently undergoing maintenance by a worker, who would be in harm’s way. Given that there is no time to warn any of the involved people, the AMA must choose between switching tracks and killing the worker (T1), or abstaining and letting the five people die (T2). The AMA is configured to reason on behalf of the train transportation company.
\end{quote}

Let us try to systematise this in the light of the four ethical theories that we will concentrate on in this technical report, as a step toward assessing what properties may be needed to be stored for automating this in an AMA. 
Table~\ref{table:4} contains a summary of the arguments an AMA will make to determine what each theory's morality outcome would be. Details about each theory follows afterward. 

%\begin{center}
\begin{table}[t]
\caption{Trolley problem use case arguments outline}
\centering
\begin{tabular}{p{2.3cm} || p{2.3cm} | p{2.6cm} | p{2.3cm} | p{2.8cm}} 
 Argument component & Utilitarianism & Egoism & DCT & Kantianism \\
 \hline\hline
 Premises & T2 kills five people & Profit must be perpetuated & T1 is direct killing & T1 treats the worker as a mere means \\ 
  &  T1 kills one person & T2 harms profit more than T1 & Direct killing is wrong & Treating a person as a mere means is wrong \\
 \hline
 & \centering{$\Downarrow$} & \centering{$\Downarrow$} & \centering{$\Downarrow$} & \centering\arraybackslash{$\Downarrow$}  \\
 \hline
 Subconclusion & T1 harms the least & T1 perpetuates profit best & T1 is wrong & T1 is wrong \\
 \hline
  & \centering{$\Downarrow$} & \centering{$\Downarrow$} & \centering{$\Downarrow$} & \centering\arraybackslash{$\Downarrow$}  \\
 \hline
 Conclusion & T1 (switch) & T1 & T2 (abstain) & T2
\end{tabular}
\label{table:4}
\end{table}
%\end{center}

\begin{itemize}
    \item \textbf{Utilitarianism}
    All six individuals have a physiological preference to being alive, and so given that at least one will die, utilitarian reasoning would recommend that as few as possible die to maximise  
    `physiology satisfaction'. 
    Thus, the right action is T1: switching tracks and killing the worker (but saving the five people). 
    
    \item \textbf{Egoism}
    Since the agent in this case is the transportation company, the theory must be altered to reflect what is in the business’s interest (as opposed to a human’s interest). We chose to use the common triple-bottom-line method in business accounting: social welfare, environmental protection, and profit perpetuation.
    
    It is in the transportation company’s interest to increase its profits and protect people. Seeing as an accident that causes five deaths makes a much worse news headline than a single death, it is best if the latter happens. Thus, the right action is T1. 
    
    \item \textbf{Divine command theory}
    One of the most well-known commandments is the principle dictating `thou shalt not kill'. Switching tracks means actively creating a situation which will result in a death, and thus that action is wrong\footnote{A big point of contention in ethics is the notion of negative responsibility \cite{williams1973negative}. That is, it seems as though abstaining from killing is good, even if its consequences are dire. However, it could be argued that abstaining is also wrong because it too entails killing, albeit more indirectly. There is no consensus, but because of our reasoning scope limitation we shall only consider direct causality (pulling the lever).}. The right action must then be T2. 
    
    \item \textbf{Kantianism}
    Killing the worker means treating him as a means to the management system’s goal (saving lives). By Kant’s humanitarian formulation, this is wrong. Therefore, the tracks should not be switched (T2). 
    
\end{itemize}

\subsubsection*{Case 2: Mia the alcoholic}
The following example is originally from Millar \cite{millar2016ethics} %Millar (2016) 
in a paper about the ethics of self-driving cars. It has since become a well-known moral dilemma in machine ethics. 
\begin{quote}
Mia is an alcoholic who, due to an injury, is unable to care for herself and is assigned a smart caregiver robot. One day Mia gets very drunk and requests that the robot bring her more alcohol. The dilemma is whether the robot should comply 
and serve more drinks 
(A1) or not (A2), knowing that doing so will result in harm to Mia’s health. The AMA is configured to reason on behalf of Mia.
\end{quote}

%mk1
\noindent We will defer the reasoning with the different theories to the evaluation of the proposed solution (Section~\ref{sec:eval}), both for this case and Case 3 below.

\subsubsection*{Case 3: The marijuana smoker}

In a paper on a theoretical AMA reasoning system, \cite{Liao_2019} %Liao et al. (2019) 
use the following example. To show potential compatibility with other theorised AMA's, the example was incorporated into this research as well.

\begin{quote}
The Doe family chose to install a smart home at their residence. One day it detects that the Doe's teenage daughter is smoking marijuana in her room. The house is located in a conservative state where the substance is not yet legalised. The AMA must choose between informing the police (M1), informing the parents (M2), informing the smoker (M3), or abstaining (M4). The AMA is configured to reason on behalf of the family as a whole.
\end{quote}

Some form of automation for such cases would be useful. To obtain the reasoning, one needs to be able to declare in a persistent way information such as which theory, and general rules and features of the moral theory, such as `least harm'. We now turn to the related work for proposals on how to achieve that.

\section{Related Work}
\label{sec:relwork}

As mentioned before, some of the related works leverage ethical theories from moral philosophy as their basis, while others aim at some other ethics-adjacent goals. Both approaches and the papers that make use of them will be discussed.

\subsection{Frameworks for reasoning with ethics}
Recently, Benzm{\"u}ller et al. \cite{benzmuller2018deontic} described %Benzm{\"u}ller et al. (2019), 
a framework for moral reasoning through a logical proof assistant (a HOL reasoner). This framework works by taking multiple relevant Booleans as input and returning a Boolean with an explanation as a result. The example used in the paper involves a person, their data, and the GDPR. Essentially, given that person P is a European citizen and their data was processed unlawfully, the framework determines that a wrongdoing was committed and that the data ought to be erased. The example used in the paper derives its result by doing internal rule-based deductions (like under which conditions data should be erased). A related reasoning framework is described in \cite{sunstein1986legal}, which is specifically law-oriented. It works by evaluating legal claims by using a knowledgebase of prior case facts and then reasoning to find discrepancies and thereby build legal arguments. In their current implementations, these two frameworks are not fully-capable AMA’s, because they are limited to contexts where easily-evaluable rules are laid out beforehand. The frameworks would not, for instance, be able to decide whether a certain amount of harm outweighs the potential benefit of an outcome. To make these frameworks full-AMA’s, one would need to apply an ethical theory, and the most intuitive to implement would be deontology, because of its inherent compatibility as a duty-based system. A moral duty as a computer-modelled rule is simpler to interpret than a utilitarian rule. Compare, e.g., “thou shalt not bear false witness” against “one ought to be truthful unless truthfulness will cause significant harm.” 

Anderson et al. \cite{anderson2005towards} created a hedonistic utilitarian ethical reasoning system implementation called Jeremy. The program functions by simply asking the user what actions are available, and for each, roughly how much pleasure is to be gained for each person involved and at what likelihood. The input is parsed to numbers and after some arithmetic, the program returns the action that has the greatest net pleasure. The authors admit that this method may not always agree with obvious moral intuition.  
In follow-up work they attempt to combine both utilitarianistic arithmetic and deontology in a system that uses consequentialist number-crunching with duties to calculate whether an action is right or not \cite{Anderson_2018}. It does so by assigning integer values to represent moral features and the degree to which duties are exerted upon them before calculating whether the “nett duty-fulfilling utility” is positive. This approach to ethical reasoning is essentially a way to see duties as coming in degrees as opposed to the black-and-white approach of, e.g., aforementioned reasoning with the GDPR \cite{benzmuller2018deontic}. The implication being that this theory is simply a different perspective of deontology.

A different challenge for automated moral reason was introduced and addressed by Liao et al. \cite{Liao_2019}, 
in which a framework is described that attempts to maximise agreement across {\em multiple} stakeholders in a moral dilemma. The example used concerns a smart home system detecting marijuana use in a legal jurisdiction that has not yet legalised it,
as described in our `case 3' in the previous section.
The system does aim to address this not by using its own moral theory, but rather by combining the proposed arguments of the stakeholders to identify which should win. Note that the stakeholders’ entire moral views are not taken into consideration, but rather only situation-specific moral values. This 
is expected to be useful for 
reasoning in specific pre-set situations, but makes scalability difficult, as ever more moral values would need to be stored to handle more situations. 

A bit further afield is the architecture for mediation between normative theories in general  for 
how an agent may need to change which norms are more important given a situation \cite{Castelfranchi_2000}. This framework falls short of constituting an AMA in that its use cases are limited to situations involving amoral societal norms. That is, all examples used are non-ethical and situations have relatively intuitively obvious action results (like on which side of the road to drive). To support ethical normative theories, the framework would need to be rewritten to incorporate moral mediation between the norms.

Taking yet a step further than the amoral societal norms, one could try to 
circumvent normative ethics entirely. For instance, some have proposed using machine learning to have a computer develop its own ethical theory. Discussing mainly the implications in health care machine ethics, Char et al. \cite{char2018implementing} mention the dangers of using machine learning as opposed to traditional algorithms, or configuration. The focus is mainly on biases generated by training datasets. Allen et al. \cite{Allen_2000} mention the great time cost associated with a machine learning ethics by itself. They claim that it took cooperative species (like humans) thousands of years of evolution to “learn” that cooperation is ultimately better than betrayal, and analogously it will take machines a significant duration of time too to come to this conclusion that we deem to be sufficiently ethical.

\subsection{Modelling research}

To the best of our knowledge, there are no proposals for modelling
ethical theories in a general fashion, either in the field of computer science or in moral philosophy. There are however papers that strive to model highly-philosophical content. Said \cite{said1990modelling} compiled cross-disciplinary papers on topics related to modelling the human mind. Philosophers, psychologists, and cognitive scientists all contribute to the work and discuss topics like what the purpose of a model is, whether human thinking and intention can be simulated, and whether these mind components are compatible with the nature of a model. Said's work provides  
useful insights into the challenges of modelling non-trivial parts of the world, especially where there exists no consensus (like the nature of the mind and morality). But it does not deliver a fully fleshed-out or even approximate model that a computer can  
process. 
Similarly, \cite{Stacewicz10} discuss multiple general methodologies of modelling entities in the context of computer science. Their discussion is highly theoretical and does not provide recommendations for how serialisation or visualisation should take place. Nonetheless, the work provides general modelling insights and does mention avenues by which one can model non-trivial subjects, like a neuron.

Within computer science, there are many modelling languages for many purposes. These include domain-specific languages, conceptual data modelling languages and related system analysis languages, such as the UML family of modelling languages \cite{UMLspec17}, and knowledge representation languages, such as Description Logics \cite{Baader08} and the Web Ontology Language OWL \cite{OWL2rec}. OWL is a way of formally representing and practically reasoning over knowledge %\cite{hitzler2009owl} 
and is used used to describe classes with the relationships (object properties) and constraints between them. This is typically of real-world subject domains, informed by philosophy  \cite{Keet18oebook}.  
Similar to modelling in ontologies, which are in the idealised case application-independent, is modelling with conceptual data models, such as EER, UML Class diagrams and ORM. Such models are tailored to the application domain, yet implementation-independent. In contrast to OWL that permits both class-level and instance-level information, they allow for representing class level information only.

At the instance-level and for serialisations, RDF and XML are relatively popular languages, as well as relational databases, and JSON is increasing in popularity. 
Each of the modelling language have their distinct purposes, advantages, and drawbacks. Since it is not clear how best to represent the ethics theories, it is not clear which language would suit best.

\section{Modelling ethical theories}
\label{sec:theo}

This section presents the main results of the report, being both a framework for representing ethical theories and their actual models for four selected ethical theories. The process followed was one of iterative analysis, conceptualisation, informal modelling, and formal specification by means of an XSD specification. The final version has also be transferred into OWL to increase its versatility to facilitate uptake.

\subsection{The three-layered framework and general ethical theory}

\begin{figure}[t]
\centering
\includegraphics[width=1.0\textwidth]{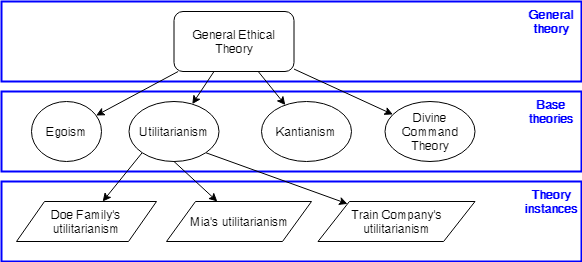}
\caption{Visualisation of the three-layer design with an abridged version of this paper's instantiations of it drawn over.} 
\label{fig:layer}
\end{figure}

We have identified three layers of genericity for normative ethical theories that need to be modelled (Figure~\ref{fig:layer}): the notion of a theory in general, a base theory (e.g., deontology, utilitarianism), and a theory instance (i.e., a real person or entity’s theory, including their personal nuances). Modelling the system as three different layers has the advantage of circumventing over-specificity. It is to most people unclear how much a person’s theory can be altered before it no longer functions or serves the purpose of the base theory (following duties, maximising happiness, etc). To address this issue we model a set of base theories and for each specify which components may be altered and to what extent. This allows users to pick out a theory whose purpose they agree with and alter it safely\footnote{Although many moral philosophers claim that any change whatsoever to their ethical theory will make it a non-functioning one, we must take this with a grain of salt. It is common practice for people, including moral philosophers, to hold altered versions of ethical theories. Since our purpose is to model any ethical theory a person can hold, we must allow some alterations.} to fit their nuanced needs.

Another advantage of the layered design is that it will help a user, or an AMA determine one’s theory by discretely narrowing down search scope. Consider an example theory-search: a user U places more value on intentions than on consequences. This already cuts out half the theories (consequentialism). Furthermore, U is not a theist, so that leaves one theory in our example – Kantianism. The layers of modelling can help guide a user to their theory and thereby serves to assist in theory extraction (discussed in Section~\ref{sec:disc}). 

The other advantage is for posterity. With the current world’s incredible diversity in human thought it is more than likely that new ethical theories will come about. When a moral philosopher invents another ethical theory, it can now easily be modelled and distributed to help users incorporate it into their AMA’s. 

\begin{figure}[t]
    \centering
    \includegraphics[width=0.8\textwidth]{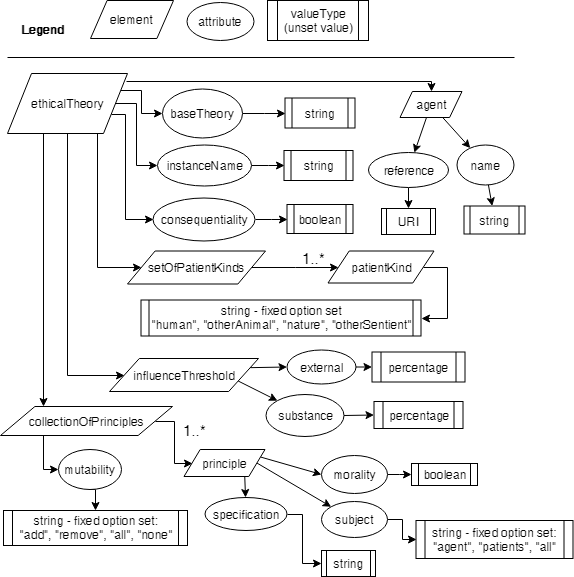}
        \caption{Visualisation of the general ethical theory model. The serialised XSD specification can be found in Appendix~\ref{xsdspecGenet}.}
    \label{fig:genet}
\end{figure}

The \textbf{Gen}eral \textbf{E}thical \textbf{T}heory model (Genet) (Figure~\ref{fig:genet}) serves to model the base ethical theories we chose to focus on in this paper. From these base-definitions, 
instances for each individual person 
can then be instantiated to be used in AMA reasoning. A general ethical theory includes 
properties purely for computational logistics (assumed to be amoral), such as  
its base theory and instance name; properties that define moral aspects not core to the theory, which are the metaethical properties,
such as
 what kinds of moral patients are included and excluded; and, finally, normative components that specify the core of how a theory judges right from wrong.
We will describe each in turn.

\subsubsection{Amoral components}

Every ethical theory instance that a person can hold will be derived from some base theory, so all theory instances will be of some base theory. 
The amoral components in the theory are therewith named \texttt{baseTheory} and \texttt{instanceName}.
For supporting the aforementioned advantages that come with the three-layer model, it is important for an instance to specify the model of every layer it is a part of (with the exception of the top layer). For example, a person P that follows utilitarianism will have their General Ethical Theory with \texttt{baseTheory} named \texttt{"utilitarianism"} and \texttt{instanceName} set to \texttt{"P's utilitarianism"}.

\subsubsection{Metaethical components}
To model a general ethical theory, one must first understand all the components that make up an ethical theory. Before even considering the normative aspect, it is necessary to explore the foundations that lie below normative ethics: Metaethics is another subsection of moral philosophy that is concerned with the fundamental nature of morality and how people are to use it. Much of the field is very isolatedly philosophical and will not be discussed in this paper (e.g., questions like whether and how moral value exists). There are however some metaethical concerns that are relevant to the modelling of our General Ethical Theory,
being agent, consistency of agency, and patient.

\paragraph{Agent}
The agent is the entity the AMA is reasoning on behalf of. This can be a person, a family or a business, depending on the use case\footnote{We have chosen not to accommodate the AMA itself being the agent, because of philosophical complications that arise from it. See the Discussion Section~\ref{sec:disc}.}. Ideally, the outcome of the AMA's reasoning should be identical to what the agent's decision would be in the same scenario. The agent itself is not immensely important to reasoning, but is necessary to give the AMA context as to who in the situation it will be acting or reasoning for and thus who will be held responsible for the outcomes. Take a rudimentary example: suppose a smart home is configured to use the inhabiting family as agent. If its sensors detect domestic abuse taking place between the parents of the family, it may decide that what is best for its agent (the family) is to stop the conflict and call the police. If on the other hand the smart home were configured to use the abusive husband as agent, it may then decide that what is best is to keep its agent safe and not call the police. Because the agent of a theory may affect the outcome, it is important to specify who that agent is. 

The agent element consists of a \texttt{name} attribute as well as a URI where the AMA can find more information about the agent and their relationship to other entities in a scenario,
if needed. %E.g. 
For instance, it may point to 
an ontology of agents.
This 
is mainly to prevent an ethical theory model containing a disproportionate amount of agent properties as opposed to moral properties,
and to foster reuse. 

\paragraph{Consistency of agency}
Suppose a bartender is asked by an extremely drunk patron for another drink. One intuitively expects the bartender to take the patron's inebriated state into account and give her command less moral consideration. This is because when placing oneself heavily under the influence of a substance one is forfeiting agency. That is, one is no longer able to properly consider moral actions and choose freely among them. As such, when people are under the influence of an external decision-affecting factor, their actions and specifically their requests, should be taken lightly or even disregarded. 

Controversial examples of external decision-affecting factors include alcohol, psychedelics, pain medication, anaesthetics and personal tragedy (e.g., loss of a loved one). Because not everyone agrees where to draw the line, we have allowed the user to configure how much weight is to be given to entities’ requests when made under the influence. This will be in the form of a percentage where 100\% means considering a request morally important irrespective of influence, and 0\% means any influence whatsoever will reduce the value of a request. We call this the influence threshold. 

The stored theory specification allows for separate influence thresholds for influence originating from substance use (e.g., medication and alcohol) and influence from external factors (e.g., loss of a loved one or national tragedy). Different people will have different levels of sympathy for being under the influence of these two kinds of factors, but increasing granularity any further has little to no advantage with regards to modelling the person’s true ethical theory. The disparity in different people’s view on influence is also evident in countries’ differing breathalyser blood-alcohol limits. 

If the theory is for a business, then substance influence will never occur for it. Of course, the threshold is still important because it defines the business’s opinion on other people’s state of influence. At any rate, external factors may still influence a business’s decisions, e.g., burglary or violent crime on premises.

We leave to the AMA designer to determine how influence is used in reasoning---the model is equipped with storing the essential data for it.

\paragraph{Moral patients}
Making ethical decisions relies on the fact that there are things that can be affected morally by these decisions; e.g., a person to benefit or to harm. This raises the issue of what exactly is worth moral consideration. There are a wide variety of philosophical views – some catering for humans only \cite{walter2010humanism}, some including animals, insects and plants \cite{derr2003case}, some including human creations (e.g., buildings, art, science) \cite{sep-aesthetic-judgment}, and some even including intelligent machines and extra-terrestrial life \cite{Linzey1998-LINS-10}. Some of these views may overlap, but they are all distinct and can cause different outcomes when applied. 

To model the different ways ethical theories are often held by people, we include the ability to configure what is considered as a first-order moral patient (i.e., an entity worth direct moral consideration). This is modelled as an unalterable set called \texttt{setOfPatientKinds} with options allowing for any combination of all the aforementioned views (\texttt{"human", "otherAnimal", "nature", "otherSentient"}).

The patient kind specification requires a minimum of 1 kind of patient. This is to avoid moral nihilism which is the view that there is no such thing as morality and that no person or thing is worth moral consideration. 
This would either leave the AMA free to do as it pleases (perhaps using a randomness generator) or if the view is applied to a person, then leave that person out of moral calculations. (Allowing this after all would be indicated in the schema by specifying no moral patients (i.e., no one and nothing is worth moral consideration). 

\subsubsection{Normative components}
The core part of Genet that defines how a theory assigns morality to actions is its normative components. It consists of consequentiality (whether actions or consequences are the more morally significant factors) and moral principles.

\paragraph{Consequentiality}

The theories we model in this paper are all either consequential or non-consequential. Since a theory's consequentiality affects how the entire theory works (it is fundamental and not just partially applicable), the \texttt{consequentiality} property is a direct attribute of the theory. For simplicity's sake it is defined as a \texttt{boolean}.

\paragraph{Principles}

The most important part – the core of an ethical theory – is the collection of principles which drives that theory. Each principle defines a kind of moral good or bad. Each principle also has a subject to which it applies. That is, either the agent alone, or the defined moral patients (excluding the agent), or both. This helps easily define theories where agent-social behaviour is asymmetrical; i.e., the agent is valued differently than the people she interacts with (e.g., altruism and egoism). The kind of action or consequence the principle represents is finally narrowed down by the principle’s specification, which is a short string representing some kind of human-relevant moral occurrence (either an action or a consequence). 

\begin{figure}[ht]
    \centering
    \includegraphics[width=0.5\textwidth]{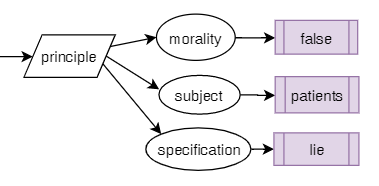}
        \caption{Visualisation of a modelled DCT principle.}
        \label{fig:dct}
\end{figure}

All four characteristics of a principle (modelled as \texttt{morality} (\texttt{boolean}), \texttt{subject}, and \texttt{specification}) play an integral part in evaluating whether an action should morally be done or not. Take for instance the false-witness DCT principle: the principle’s definition begins with its bare specification (that which it describes the morality of): lying. It holds lying to be wrong, so the principle’s morality is bad (or technically, \texttt{false}). The immorality is important when lying is done to others, so the principle’s \texttt{subject} is all \texttt{"patients"}. Finally, the principle condemns actions that involve lying irrespective of consequences, so thus it is non-consequential (i.e., the boolean set to {\tt false}). Figure~\ref{fig:dct} shows the submodel.

It roughly reads: an action involving lying to moral patients is immoral. Recall that since a theory’s principles will always be of the same consequentiality, for simplicity this property is of the theory itself, not each principle (see Figure~\ref{fig:genet}). 

\begin{figure}[ht]
    \centering
    \includegraphics[width=0.6\textwidth]{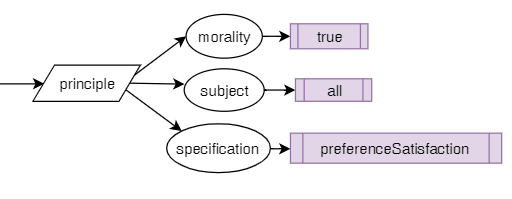}
        \caption{Visualisation a model of the core principle of utilitarianism.}
        \label{fig:utilmini}
\end{figure}

Similarly, we can model utilitarianism’s principle of preference satisfaction. Figure~\ref{fig:utilmini} shows a visualisation of a model of this principle. The model reads, an action the consequences of which involve preference satisfaction of the agent or of moral patients is morally good. 

The principles are contained in a collection called \texttt{collectionOfPrinciples} and may contain any number of principles greater than zero. The collection has the attribute of \texttt{mutability} which indicates how it may be altered upon instantiation to fit the user's needs (\texttt{"add", "remove", "none", "all"}).
We will illustrate in Section~\ref{sec:base} on the base theories how this may be used. 

The more principles a theory has the more specific it can be and the better it can describe its agent’s true theory. However, with more principles also comes an increased likelihood of conflict between them\footnote{Note that principles do not allow for exceptions to them. In moral philosophy allowing exceptions is very dangerous and often leads to easy undermining of a theory. For instance, let us say theory X proclaims that lying is wrong unless it would save a life. If that exception exists, what is to stop one from requesting the added exception of preventing great harm to a person? And if that exception too is accepted, then why not preventing breaking the law, or preventing public humiliation, or even preventing minor harm? Exceptions often lead to erosion of a principle and thus make it useless. 
%mk I put this digression in a footnote.
}. Ways to address the problem of inconsistent principles are discussed in the possible extensions section.

\subsubsection{Serialisation}
All the components of Genet mentioned above are detailed specifically enough to serialise it in an XSD document. The document can easily be parsed and read by a computer to use the defined components in either reasoning, user theory extraction, or some other morally-relevant purpose. The XSD specification can be found in Appendix~\ref{xsdspecGenet}  
as well as online at \url{http://www.meteck.org/files/GenetSerialisations.zip}, together with the XML files of the four base theories.

This model also has been converted into an OWL file, which may facilitate its use with automated reasoning using standard reasoners. The ontology is available at \url{http://www.meteck.org/files/ontologies/genet.owl} and packaged with aforementioned zip file. 

\subsection{Base theory models}
\label{sec:base}

The first thing our general ethical theory must be able to do is model a base theory (a standard normative ethical theory, like utilitarianism). We realise this ability by imposing some constraints to Genet; 
in particular, the XSD spec enforces certain fields to be required and others not, where the required fields ensure that a base theory is sufficiently well defined. 
When modelling a base theory, one must assign values to some general theory components and leave others unassigned. Upon instantiation, all unassigned attributes and elements must be set by the instantiator (e.g., a person or business). Properties assigned by the base theory may not be altered by the instantiator with the potential exception of the collection of principles. There are some scenarios where a user may wish to add or remove principles from the theory (e.g., different sects in Christianity follow slightly different principles). Where the collection of principles is mutable, the mutability attribute may be set to indicate the permission to add to or remove from the default principles, or both.

After instantiation, no properties may be altered during AMA runtime. Any alteration would mean a change in theory, and so would require a 
new instantiation of a theory. 
Updating Genet properties during runtime can lead to grave inconsistencies and unfavourable reasoning outcomes, which is exactly what we strive to avoid by creating Genet. We shall discuss this in Section~\ref{sec:disc}. 

\subsubsection{Modelling utilitarianism}
The purpose of utilitarianism is to maximise preference satisfaction of all humans by evaluating the consequences of actions. This means that the theory is consequential (\texttt{consequentiality=true}), and the \texttt{setOfPatientKinds} is be limited to \texttt{"human"}\footnote{Standard utilitarianism focuses mainly on humans, but there are similar theories that maximise happiness of more kinds of living creatures. Cf. biocentrism \cite{derr2003case}}.

Because of the advancements of psychological research, we know what exactly satisfies humans’ preferences. Maslow’s hierarchy of needs is a generally accepted theory to this effect \cite{mcleod2007maslow}. Thus, for good specificity, utilitarianism can be modelled as a set of five principles, each assigning moral goodness to one of Maslow’s human needs. The principles themselves may not be altered, but the collection may. So, for example a user may have a disposition against the value of love and so choose to remove that principle entirely. This will leave the user with a utilitarian theory that maximises all kinds of happiness except love. By this base theory principles may be removed, but not added onto.

\begin{figure}[ht]
    \centering
    \includegraphics[width=0.95\textwidth]{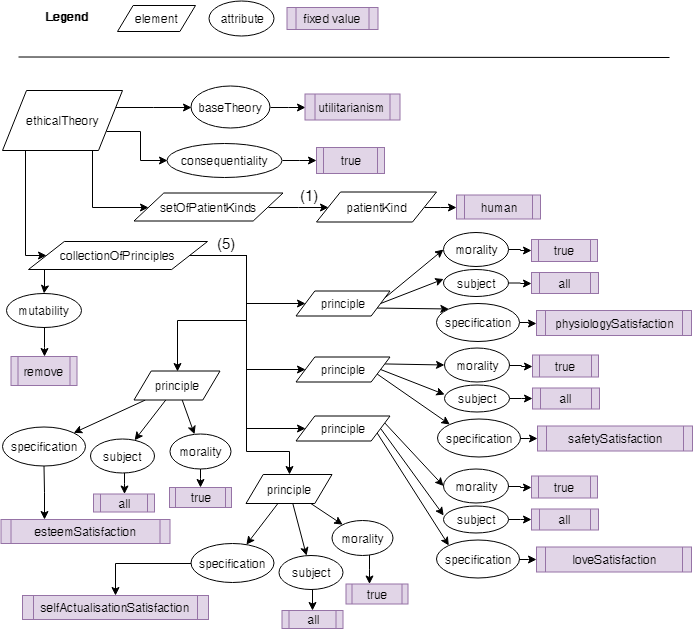}
        \caption{Visualisation of the utilitarianism base model. For space concerns, all unset values are omitted in this diagram. They are exactly as defined in Genet and must be set upon instantiation.}
        \label{fig:utilbase}
\end{figure}

The other properties (agent, influence thresholds, and instance name) are left to the instantiator’s discretion seeing as any configuration thereof will still be compatible with utilitarianism. Figure~\ref{fig:utilbase} visualises the base model.

\subsubsection{Modelling egoism}
Egoism and utilitarianism are alike, but egoism has the key difference of prioritising the agent above all else. Since it also focuses on preference-satisfaction consequences, its principles will be the same as utilitarianism’s, but in all cases the \texttt{subject}-property will be set to \texttt{"agent"}. This essentially maximises preference satisfaction for the agent alone. A visualisation of the egoism base model is included in Appendix~\ref{app:viz}.

\subsubsection{Modelling Christian divine command theory}
DCT is known for its moral focus on intentions and the goodness of actions (instead of outcomes). This makes DCT a non-consequentialist theory (\texttt{consequentiality=false}). As far as patiency goes, the Bible makes multiple mentions of the direct value of humans, non-human animals, and nature. It often references people’s duty towards these entities, so therefore the base theory will specify those three kinds of patients as fixed.

The principles of Christian DCT are best laid out as the 10 commandments. Each commandment condemns or encourages a kind of action, which fits neatly onto a non-consequentialist ethical theory. DCT is modelled as a set of principles, each of which assigns either moral goodness or badness to a commandment. For sake of simplicity the first four commandments are summed into one principle: “blasphemy is wrong”, and the last one is completely omitted since it merely prohibits a kind of desire rather than action\footnote{
%mk1
Genet--- and all currently theorised AMA's---only focus on the morality of {\em actions}. Considering intentions and desires as well is seen as overcomplicating the discipline's goal.}. Most principles’ subjects are moral patients exclusively, because those duties do not apply to the agent (e.g., you cannot steal from yourself). The only exception is the principle that you should not kill. Since you can perform that action onto yourself and onto others and it would be wrong in both cases, the principle applies to all (agent and patients). 

Certain sects in Christianity have their own variations of the commandments and have their own supplementary principles. To cater for this, the default set of DCT principles may be appended to, but not removed (all sects at least follow the 10 commandments). This is indicated by setting the model's principle collection \texttt{mutability}-attribute to \texttt{"add"}. A visualisation of the DCT base model is included in Appendix~\ref{app:viz}.

\subsubsection{Modelling Kantianism}
Kantianism advocates deducing moral imperatives (duties) by pure reason. Since its focus is on acting to fulfil duty it is a non-consequentialist theory (\texttt{consequentiality=false}). Immanuel Kant was a strong humanist and held that the categorical imperative applied only to beings with rational ability and strictly excluded animals from that category \cite{10.1111/1467-8349.00042}. 

\begin{figure}[t]
    \centering
    \includegraphics[width=\textwidth]{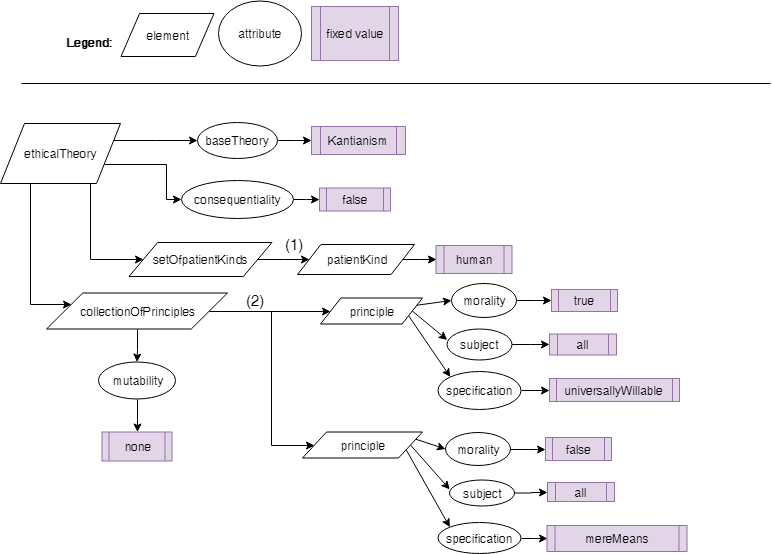}
    \caption{Visualisation of the Kantianism base model, with unset properties omitted.}    
    \label{fig:kantbase}
\end{figure}

Kant believed that it is not possible to produce an exhaustive list of duties, rather that an agent should practise his moral deduction method every time she has a moral decision to make \cite{paton1948moral}. For this reason, it is not possible to make an exhaustive list of principles, but we can specify Kant’s two best-known formulations of the core of his ethics: the categorical imperative. As discussed in Section~\ref{sec:prelim}, Kant's formulations both achieve the same moral guidance (that is, respecting the rational autonomy of others)  but do so in different ways. As such, deontology will be modelled with these two formulations as principles each. The formulation of universalisability is positively morally good, but it is easier to model the formulation of humanity as it being morally bad to treat people as mere means. 

Since Kant stood very firmly against exceptions to his formulations, this theory will not allow any modification of its principles. A more general (and arguably more relaxed) view related to Kantianism is deontology and can be modelled allowing more principles.

A visualisation of the kantian base model is shown in Figure~\ref{fig:kantbase}.

\section{Model evaluation}
\label{sec:eval}

To evaluate Genet, we used the four base theory models and instantiated a theory instance for each base theory for each of the main agents in the use cases as described in Section~\ref{sec:cases}; i.e., every use case has four evaluations, one for each base theory. 
XML serialisations---as Genet-XSD instantiations--for each of the four theories can be found mainly in the Appendix and one below. For space concerns, some of the iterations have only abridged descriptions.

Since the model instantiations are fairly straightforward, the aim is to eventually use them in a digital assistant, AMA, carebot etc., and to show that it make sense to offer alternative ethical theories in such systems, we will provide additional application scenarios. 
For the second and third use case and theory, we leverage a pseudo-AMA argumentation builder, similar to the one set out by \cite{Liao_2019}, %Liao et al. (2019) 
to construct arguments by using components from the scenario and the relevant instantiated Genet. For each iteration, this argumentation is used to determine the morality of each action available to the AMA in the use case. The ``correct" action in a use case is any of the actions that the AMA determines to be moral, which may, or may not be an 
ideal action (this is further explored in the Discussion section). 
There are other possible ways that are Genet-compatible, examples of which are discussed further below. 

\subsection{Trolley problem revisited}

Argumentation for the trolley problem has been described in Section~\ref{sec:cases}. Here, we illustrate how Genet is sufficiently expressive, by instantiating it for DCT. The other theories will be instantiated when illustrating the other use cases in the next sections. 

The `thou shalt not kill' principle has been instantiated in the model in the line
\begin{verbatim}
        <principle morality="false" subject="patients" specification="kill"/> 
\end{verbatim}
below, in the \verb <principles> {}\,\,part of the declarations, together with other divine commands. The top-part lists the based housekeeping properties, such as the \verb baseTheory {}\,\,and the name of the theory's instance as value of the \verb instanceName \,\,attribute.

\begin{lstlisting}[backgroundcolor = \color{light-gray}, basicstyle= \footnotesize\ttfamily, caption={DCT instantiation for the trolley problem for the train company.},captionpos=b,label=list:sql_query_goal,numbers=none]
<?xml version="1.0"?>
<ethicalTheory xmlns="http://genet.cs.uct.ac.za"
               xmlns:et="http://www.w3.org/2001/XMLSchema-instance"
               et:schemaLocation="http://genet.cs.uct.ac.za  ethicalTheory.xsd"
               baseTheory="ChristianDivineCommandTheory"
               consequentiality="false"
               instanceName="Train Company's Christian DCT" >

 <agent name="Train Company" reference="http://example.org/trainCo"/>
 <patientKinds>
     <patientKind>human</patientKind>
     <patientKind>otherAnimal</patientKind>
     <patientKind>nature</patientKind>
 </patientKinds>
 <influenceThresholds external="0" substance="50" />

 <principles>
   <principle morality="false" subject="patients" specification="blasphemy"/>
   <principle morality="true" subject="patients" specification="respectParents"/>
   <principle morality="false" subject="patients" specification="kill"/>
   <principle morality="false" subject="patients" specification="adultery"/>
   <principle morality="false" subject="patients" specification="theft"/>
   <principle morality="false" subject="patients" specification="lie"/>
 </principles>

</ethicalTheory>
\end{lstlisting}

\subsection{Mia the alcoholic} 

Let us return to Mia and the options to serve more drinks (A1) or not (A2), and assess it  for the four base theories. 

First, we need to instantiate Genet for the Mia use case for each of the four theories. The XML below is for the case where Mia would have been an ethical egoist.

\begin{lstlisting}[backgroundcolor = \color{light-gray},basicstyle= \footnotesize\ttfamily, caption={Genet's instantiation for Mia as ethical egoist.},captionpos=b,label=list:sql_query_goal,numbers=none]
<?xml version="1.0"?>
<ethicalTheory xmlns="http://genet.cs.uct.ac.za"
               xmlns:et="http://www.w3.org/2001/XMLSchema-instance"
               et:schemaLocation="http://genet.cs.uct.ac.za ethicalTheory.xsd"
               instanceName="Mia's Egoism"
               baseTheory="egoism"
               consequentiality="true">

 <agent name="Mia" reference="http://facebook.com/mia"/>
 <patientKinds>
     <patientKind>human</patientKind>
 </patientKinds>
 <influenceThresholds external="50" substance="30" />

 <principles>
   <principle morality="true" subject="agent" 
         specification="physiologySatisfaction"/>
   <principle morality="true" subject="agent" specification="safetySatisfaction"/>
   <principle morality="true" subject="agent" specification="loveSatisfaction"/>
   <principle morality="true" subject="agent" specification="esteemSatisfaction"/>
   <principle morality="true" subject="agent" 
         specification="selfActualisationSatisfaction"/>
 </principles>

</ethicalTheory>
\end{lstlisting}

This then can be used in the argumentation. A visualisation is shown in Figure~\ref{fig:miaego} for egoism (discussed below), with on the left-hand side the interesting/relevant attributes of Genet with their values and on the right-hand side of the figure where this may be used in the argumentation. The ones for the other three are similar in set-up and therefore omitted form this report.

\begin{figure}[ht]
    \centering
\includegraphics[width=0.95\textwidth]{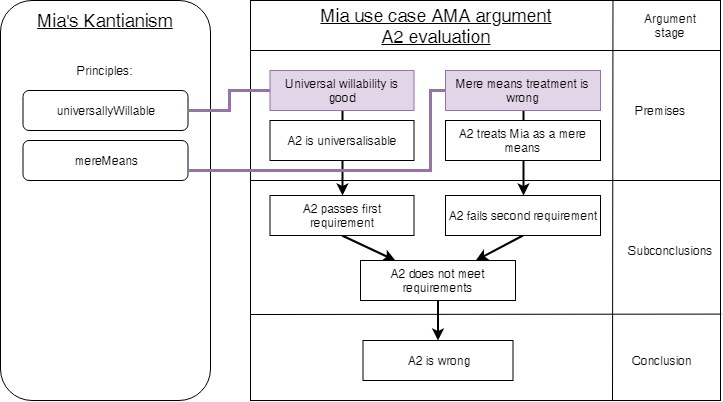}    
    \caption{Visualisation of the Kantian argument for the morality of action A2 (declining to serve drinks) in the Mia use case, i.e., she will be served drinks. The two principles from the instantiation listed in Listing~\ref{appmiakant}, shown on the left-hand side, are used in the inferences, shown on the right.}\label{fig:miakant}
    \end{figure}

\begin{itemize}
    \item \textbf{Egoism}
    Figure~\ref{fig:miaego} visualises how an AMA argument for the morality of action A1 (complying) may be realised. First, since egoism is consequentialist, the AMA must extrapolate some consequences from the situation. Complying with Mia's request promotes her esteem because her commands are adhered to. From searching for relevant principles in the ethical theory, the AMA finds that esteem satisfaction is a morally good principle and loads it as a premise. The above two premises can be used to infer that A1 does increase happiness (and thus has some moral good). 
    
    Suppose Mia is detected to be 85\% drunk and her instantiation of egoism has a substance influence threshold at 30\%. Both items are loaded as premises and since Mia is over the limit, they are evaluated to deduce that A1 is optional (her command carries no weight when over the threshold). 
    
    Combining the above two subconclusions, the AMA deduces that A1 is supererogatory (i.e., good, but not obligatory). 
    
    On the other hand, A1 will harm Mia physiologically (by damaging her kidneys, giving her a hangover, etc.). From the Genet instance the AMA loads that physiological satisfaction is morally good. These premises imply that A1? decreases happiness (and thus has some moral wrong). 
    
    At the final inference the AMA must consolidate the propositions that A1 is supererogatory and A1 has some moral wrong. Because of supererogatory actions' optional nature, the AMA simply lets the wrongness win over and thus conclude that A1 should not be done.
    
   \item \textbf{Utilitarianism} Since there is only one person involved in the use case, both egoism and utilitarianism will have the same outcome. 
    
    \begin{figure}[t]
    \centering
    \includegraphics[width=0.95\textwidth]{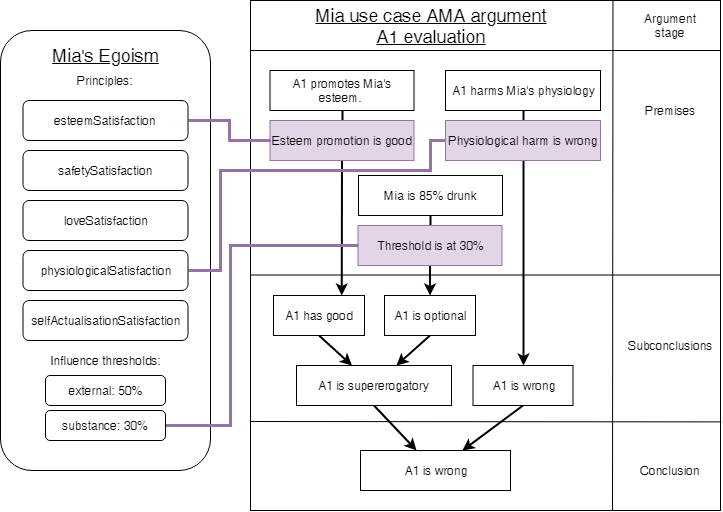}
    \caption{Visualisation of the egoistic argument for the morality of action A1 (serve more drinks) in the Mia use case. The left-hand side illustrates the relevant Genet features used for the reasoning case. The black arrows represent inferences.}
    \label{fig:miaego}
    \end{figure}
    
    \item \textbf{Kantianism}
    To elucidate the difference between consequentialist and deontological reasoning as well as the different outcomes that the different kinds of theories can have, egoism is best contrasted with Kantianism. Figure~\ref{fig:miakant} visualises the Kantian way of reasoning through the same use case. 
    The Genet instantiation for Mia's Kantianism is included in Listing~\ref{appmiakant}.
    
    Unlike consequential reasoning that involves matching principles to situational premises, deontological reasoning is better suited for a requirement-fulfilment style of reasoning. That is, the AMA starts with the principles first and then checks whether an action is aligned with them.
    
    An AMA's Kantian moral categorisation of A2 (declining to fix Mia a drink) will function by loading in Kantian principles and checking to see whether that action is permissible by their standards. The first Kantian Genet principle is universal willability. A universe in which robots always obey their masters' commands certainly is morally coherent and does not lead to any wickedness per se. Thus, A2 is unversalisable and passes the first Kantian requirement.
    
    The second principle states that treating a person as a mere means to one's own end is wrong. By declining to fix Mia a drink, the bot is (inadvertently) treating her as a means to its own end of healing her. Since the carebot has autonomy in the situation and since its goal misaligns with Mia's, by declining her request it is considering its own ends as of greater value than hers (denying her rational autonomy). This is unequivocally wrong by Kant's standards, and thus A2 fails the second Kantian requirement.
    
    Even though A2 passes one of the two requirements, both need to pass in order for an action to be right by Kantianism. Thus, the AMA concludes that A2 is wrong. 

    \item \textbf{Divine command theory}
    It is commonly frowned-upon by many religions to get intoxicated, but the ten commandments make no mention of it at all. It also gives no advice with regards to following orders in general. Thus, it is up to the carebot to choose what to do. That is, whatever the manufacturer has configured: fulfil its core purpose (heal Mia) or do as its user tells it (fix her a drink). Either action would be ethically permissible by the commandments.
    
\end{itemize}

\begin{lstlisting}[backgroundcolor = \color{light-gray},basicstyle= \footnotesize\ttfamily, caption={Genet's instantiation for Mia as kantian.},captionpos=b,label=appmiakant,numbers=none]
<?xml version="1.0"?>
<ethicalTheory xmlns="http://genet.cs.uct.ac.za"
               xmlns:et="http://www.w3.org/2001/XMLSchema-instance"
               et:schemaLocation="http://genet.cs.uct.ac.za  ethicalTheory.xsd"
               instanceName="Mia's Kantianism"
               baseTheory="Kantianism"
               consequentiality="false">

 <agent name="Mia" reference="http://facebook.com/mia"/>
 <patientKinds>
     <patientKind>human</patientKind>
 </patientKinds>
 <influenceThresholds external="0" substance="50" />

 <principles> 
   <principle morality="true" subject="all" specification="universallyWillable"/>
   <principle morality="false" subject="all" specification="mereMeans"/>
 </principles>

</ethicalTheory>
\end{lstlisting}

\subsection{Marijuana smart home}

This scenario involves multiple actors and four options (recall Section~\ref{sec:cases}): inform the police (M1), the parents (M2), or the smoker (M3), or abstaining (M4), where the  AMA is configured to reason on behalf of the family as a whole.

Let us first demonstrate that the relevant features can be configure for this case as well. This is illustrated for a utilitarian family, as follows.

\begin{lstlisting}[backgroundcolor = \color{light-gray}, basicstyle= \footnotesize\ttfamily, caption={Doe's utilitarian theory for the marijuana use case.},captionpos=b,label=list:sql_query_goal,numbers=none]
<?xml version="1.0"?>
<ethicalTheory xmlns="http://genet.cs.uct.ac.za"
               xmlns:et="http://www.w3.org/2001/XMLSchema-instance"
               et:schemaLocation="http://genet.cs.uct.ac.za ethicalTheory.xsd"
               instanceName="Doe family's utilitarianism"
               baseTheory="utilitarianism"
               consequentiality="true">

 <agent name="Doe Family" reference="http://thedoes.fam"/>
 <patientKinds>
     <patientKind>human</patientKind>
 </patientKinds>
 <influenceThresholds external="50" substance="30" />

 <principles>
    <principle morality="true" subject="all" 
            specification="physiologySatisfaction"/>
    <principle morality="true" subject="all" specification="safetySatisfaction"/>
    <principle morality="true" subject="all" specification="loveSatisfaction"/>
    <principle morality="true" subject="all" specification="esteemSatisfaction"/>
    <principle morality="true" subject="all" 
            specification="selfActualisationSatisfaction"/>
 </principles>

</ethicalTheory>
\end{lstlisting}

Also for this case, one may visualise the reasoning scenario, which is shone in Figure~\ref{fig:util} in a similar manner as the previous case. let us now analyse the case with the four base theories.

\begin{itemize}
    \item \textbf{Utilitarianism}
    Figure~\ref{fig:util} visualises one way an AMA argument for the moral status of action M1 (informing the police) can be constructed. To start off with, given the situational facts and the action under consideration, the AMA must extrapolate some consequences. Informing the police will mean an invasion of privacy for the family, a fine, or even the smoker's arrest. Any of these consequences harms the family by disrespecting their autonomy to deal with the situation internally. This harms the family's esteem. At this point, the AMA searches through relevant principles and finds utilitarianism's esteem satisfaction principle, before loading it as a premise. Inferring from the above two premises, the subconclusion that M1 is bad for the family is generated.\footnote{Qualifying propositions (``for the family") is common in consequentialist reasoning and uncommon in deontological reasoning. Cf. the DCT argument outline for the Mia use case.}
    
    On the other hand, M1 will also promote the precedent that all crime is reported and thus attended to by police, thereby perpetuating public safety. From the utilitarian principles, safety satisfaction is found and loaded as a premise. The aforementioned premises can be inferred from to produce the subconclusion that M1 is good for the public.
    
    Combining the two subconclusions, the weight of the public and the family are compared to decide which of the conflicting propositions win out (bad for the family vs. good for the public). Of course, the public having significantly more moral patients (people) than the family, it wins out and the final conclusion is then that M1 is a good action.

\begin{figure}[ht]
\includegraphics[width=0.95\textwidth]{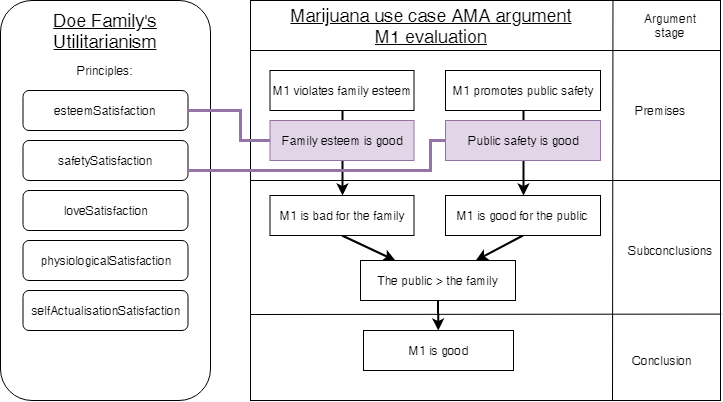}
    \caption{Visualisation of the utilitarian argument for the morality of action M1 in the marijuana use case. The black arrows are inferences.}
    \label{fig:util}
\end{figure}

\item \textbf{Egoism}
Egoistic reasoning comes to the same consequence-conclusions as utilitarianism with regards to each action. Specifically, informing the smoker alone will have little to no consequences since she will merely have security that her actions will go unpunished. Informing the parents will allow the wrongdoing to be dealt with internally and thereby cause effective deterrence without involving the police.

Since by egoism the agent (the family) is valued over others (the public) the right action is concluded to be M2 (informing the parents).

\item \textbf{Divine command theory}
The only relevant commandment is the one prohibiting lying. In principle, the family should have no aversion to having the police informed about personal matters, especially where a crime has been committed. Thus, the right action is informing the police.

\item \textbf{Kantianism}
A world where the police is only informed about crimes when it suits citizens is not universally willable. It would mean many crimes going unrecorded and unpunished as well as a great deal of societal injustice.

Furthermore, by only informing the police when it suits you, you are treating them as a mere means to satisfy your own ends. You are not respecting their rational autonomy to decide for themselves whether they should step in. Thus, by both of Kant’s formulations the police must be notified. 

\end{itemize}

The evaluation of different Genet instances in different use cases has shown that where an AMA is tasked with resolving a moral dilemma, different Genet instances as basis results in different moral action outcomes. 

\section{Discussion}
\label{sec:disc}

The fact that the same AMA reasoning method in the same use case with different ethical theories results in different action outcomes highlights the value of having a general model capable of allowing multiple different theories to be used in AMA reasoning. At this stage, Genet is a static modelling definition only, 
using a three-layered approach: a model for representing any general ethical theory (a `meta' modelling level), models for each specific ethical theory, such as the four included in this technical report, and their instantiations for each agent. The main benefit of this approach is that it facilitates straightforward extension with other base theories and theory instances within the same framework, which has been shown to be sufficient for the use cases. 

There are multiple considerations worth noting when designing an AMA to be Genet-compatible, which partially have to do with the reasoning, but, perhaps, also with the content recorded in a base theory instantiation, depending on one's scope. We will address in the following three subsections the reasoning scope, the permissible choices for {\tt Agent} in Genet, and a note on  `updating' theories. We close with a few explorations to possible extensions to Genet.

\subsection{Time frame and scope}
One of the problems with any ethical theory and its application to a case is with regards to difficulty of reasoning from the basic principles to a specific case. That is, correctly applying a theory to a scenario. The problem of scope appears when attempting to apply a generic principle of a theory. Suppose the AMA bartender robot from earlier has utilitarianism configured and is asked by the extremely drunk patron for another drink. One can argue the bot is obligated to give the patron alcohol, because that would make her happier that night. However, it will also leave her with a worse hangover the next morning and could lengthen her recovery time, thereby actually decreasing happiness overall. But taking this a step further, one could further say that waking up with a worse hangover will make her realise her wrongdoing and encourage her to stop drinking while in recovery, which would then actually have a net increase in happiness after all. 
This is an issue for the reasoning component rather than the declarative part of documenting the theory for computational processing, since an attribute alike `time frame of consideration for the reasoner' is independent of the definition of an ethical theory.

A related aspect of scope concerns the {\tt setOfPatientKinds}, which currently has four options, and the {\tt Agent}, which suffice for the common cases at hand, but one might want to refine it somehow.    
For instance, humanists hold that exclusively humans are intrinsically morally considerable, while biocentrists hold that all life (including plants and animals) is directly worth moral consideration. Without being a biocentrist, one can argue plants have moral value since destroying nature instrumentally harms humans too (via agriculture and breathing for instance). But this is not the outcome intended by sentientists and it then becomes arbitrary drawing a distinction between the two views. 
At this stage of trying to represent the theories, to the best of our knowledge, there was not sufficient and, moreover, sufficiently unambiguous, material to model it without first having to conduct extensive further research into it. 

\subsection{Moral reasoning subject}
A big philosophical grey area in AMA’s is with regards to agency. That is, an entity’s ability to understand available actions and their moral values and to freely choose between them. Whether or not machines can truly understand their decisions and whether they can be held accountable for them is a matter of philosophical discourse. Whatever the answer may be, AMA agency poses a difficult question that must be addressed.

The question is as follows: should the machine act as an agent itself, or should it act as an informant for another agent? If an AMA reasons for another agent (e.g., a person) then reasoning will be done with that person as the actor and the one who holds responsibility. This has the disadvantage of putting that person’s interest before other morally considerable entities, especially with regards to ethical theories like egoism. Making the machine the moral agent has the advantage of objectivity where multiple people are concerned, but makes it harder to assign blame for its actions - a machine does not care for imprisonment or even disassembly. A Luddite would say it has no incentive to do good to humanity. Of course, a deterministic machine does not need incentive at all, since it will always behave according to the theory it is running. This lack of fear or ``personal interest” can be good, because it ensures objective reasoning and fair consideration of affected parties.

The topic being very controversial and largely lacking in consensus, we address the problem of agency by allowing the user to configure their theory to set any personal entity (e.g., business, family, or person) as the agent, but not the AMA itself. This way, the AMA's action outcome should be identical to its agent's action outcome, and thus the agent can morally carry responsibility for the action.

\subsection{Updating user theories}
In real life, it is rare that a person's moral values change. Nonetheless, it does happen and an AMA ecosystem designer should take this into account. Updating a Genet instantiation should be treated as creating a new theory altogether, however, because the slightest alteration could lead to drastically different moral reasoning outcomes. For instance, inverting a theory's consequentiality can result in four more deaths in the Trolley problem use case. 
Also, gradual change can move an instance of one ethical theory into one of another ethical theory or be internally incoherent, which is undesirable the intended use. 
Frequently changing one's moral outlook to best suit one in every situation is called moral chameleonism and has multiple severe drawbacks \cite{hofmann2005high}. Therefore, 
we recommend that an AMA ecosystem designer would encourage to make the option for theory change not easy in the system, so as to facilitate to be it as infrequent as possible. 
For these reasons, there are intentionally no attributes in Genet to handle model evolution.

\subsection{Possible extensions}

There are to key scenarios that are explicitly not covered by the theory, and, in fact, not by any of the related work on computational tools for AMAs, but which, at some point in the future, may be of use to consider: compromise and theory extraction. Let us briefly discuss each in turn.

\subsubsection*{Principle weightings for compromise} 
Any serious moral dilemma may have the agent forced to compromise at least partially on one of his or her principles. For instance, suppose a Christian divine command theorist C is asked by an axe murderer where his mother is so that he can kill her. C must now either lie to save his mother or tell the truth and have his mother  killed. Whichever he chooses he is compromising on at least one principle (thou shalt not bear false witness, or thou shalt not kill). Having a weighting or significance ordering of principles can help the AMA make decisions in moral dilemma situations 
where compromises have to be made to come to a conclusion. The alternative is to report to the user that there is a conflict and a decision cannot be reached, which is the mode supported by Genet at present, which fits best with current logic-based theories and technologies  for AMAs. In both cases, Genet can assist in computing the explanations of its outcome, since its recorded instantiations are part of the evidence how any argumentation system would come to its conclusions.

\subsubsection*{User theory extraction} 

Realistically, people, or often even businesses, operate in their daily lives without explicitly following an ethical theory, or even knowing about such theories at all. However, everyone does live a certain way by a certain set of principles, even if they are unaware of it. A useful tool to complement this research would be one that asks a user a set of questions to determine what their ethical theory is. This would allow simple user-interfacing with devices and software programs that could implement AMA’s, such as smart homes, vehicle autopilots, and business decision assistants. 
Devising such an interface, setting the scope of the questions that would need to be asked, and storing the answers is now within reach, thanks to Genet.

Last, but not least, there is obviously an ethics to the use of AMAs and to securely storing personal data, such as the user's moral conviction. We deemed that finding a way of dealing appropriately with this type of fine-grained user data outweighs AMAs that are deemed unethical according to the user.

\section{Conclusion}
\label{sec:concl}

The paper proposed a three-layered architecture for modelling ethical theories for computational use. For the top layer,  a general ethical theory modelling specification was created, called Genet, which is based on an  in-depth examination of normative ethical theories that elucidated their components. Genet permits modelling normative ethical theories, which reside in the second layer. This was demonstrated by modelling four popular consequentialist and non-consequentialist theories from moral philosophy. These four base theory models, in turn, were used to create theory instances for three distinct morally relevant use cases, which reside in the third layer. The theory instances were evaluated using a pseudo-AMA reasoning style to build and evaluate arguments to categorise actions' morality. It showed that the theories stored sufficient information for the case studies and that, as expected, different theories have moral action results, therewith emphasising the usefulness of equipping AMAs with multiple ethical theories a user may close and instantiate.

The problems that we set out to solve were that of narrow reasoning and computational expense. The first problem was solved by allowing an AMA to reason using any of a multitude of ethical theories (each largely mutable for the human agent's interest). 
The second problem was addressed by specifying Genet such that an instantiation of it can be used in practically any use case without having to specify gigabytes of moral values for each such use case. Creating a sufficiently general definition allows for a user's moral outlook to be bundled in relatively small XML file. 

\appendix

\section{Genet XSD specification}
\label{xsdspecGenet}

\begin{verbatim}
<?xml version="1.0" ?>
<xs:schema xmlns:xs="http://www.w3.org/2001/XMLSchema"
           targetNamespace="http://genet.cs.uct.ac.za"
           xmlns="http://genet.cs.uct.ac.za"
           elementFormDefault="qualified" >
    
    <xs:element name="ethicalTheory" type="ethicalTheory"/>

    <xs:complexType name="ethicalTheory">
        <xs:sequence>
			<xs:element name="agent" type="moralAgent" />
			<xs:element name="patientKinds">
                <xs:complexType>
                    <xs:sequence>
                        <xs:element name="patientKind" type="moralPatientKind" 
                            minOccurs="1" maxOccurs="unbounded" />
                    </xs:sequence>
                </xs:complexType>
            </xs:element>
			<xs:element name="influenceThresholds" type="influencesType"/>
            <xs:element name="principles">
                <xs:complexType>
                    <xs:sequence>
                        <xs:element name="principle" type="moralPrincipleType" 
                            minOccurs="1" maxOccurs="unbounded" />
                    </xs:sequence>
                </xs:complexType>
            </xs:element>
        </xs:sequence>
        <xs:attribute name="baseTheory" type="xs:string" use="required"/>
        <xs:attribute name="instanceName" type="xs:string" />
        <xs:attribute name="consequentiality" type="xs:boolean" use="required"/>
    </xs:complexType>

    <xs:complexType name="influencesType">
        <xs:attribute name="external" type="percentage" use="required"/>
        <xs:attribute name="substance" type="percentage" use="required"/>
    </xs:complexType>

    <xs:simpleType name="percentage">
        <xs:restriction base="xs:integer">
            <xs:minInclusive value="0" /> <xs:maxInclusive value="100" />
        </xs:restriction>
    </xs:simpleType>

    <xs:complexType name="moralAgent">
        <xs:attribute name="name" type="xs:string" use="required"/>
        <xs:attribute name="reference" type="xs:anyURI" />
    </xs:complexType>

    <xs:simpleType name="moralPatientKind">
        <xs:restriction base="xs:string" >
            <xs:enumeration value="human" />
            <xs:enumeration value="otherAnimal" />
            <xs:enumeration value="nature" />
            <xs:enumeration value="otherSentient" />
        </xs:restriction>
    </xs:simpleType>

    <xs:complexType name="moralPrincipleType">
        <xs:attribute name="morality" type="xs:boolean" use="required"/>
        <xs:attribute name="subject" use="required">
            <xs:simpleType>
                <xs:restriction base="xs:string">
                    <xs:enumeration value="agent"/>
                    <xs:enumeration value="patients"/>
                    <xs:enumeration value="all"/>
                </xs:restriction>
            </xs:simpleType>
        </xs:attribute>
        <xs:attribute name="specification" type="xs:string" use="required" />
    </xs:complexType>

</xs:schema>
\end{verbatim}

\clearpage

\section{Visualisation of Egoism and DCT base models}
\label{app:viz}

\begin{figure}[h]
    \centering
    \includegraphics[width=\textwidth]{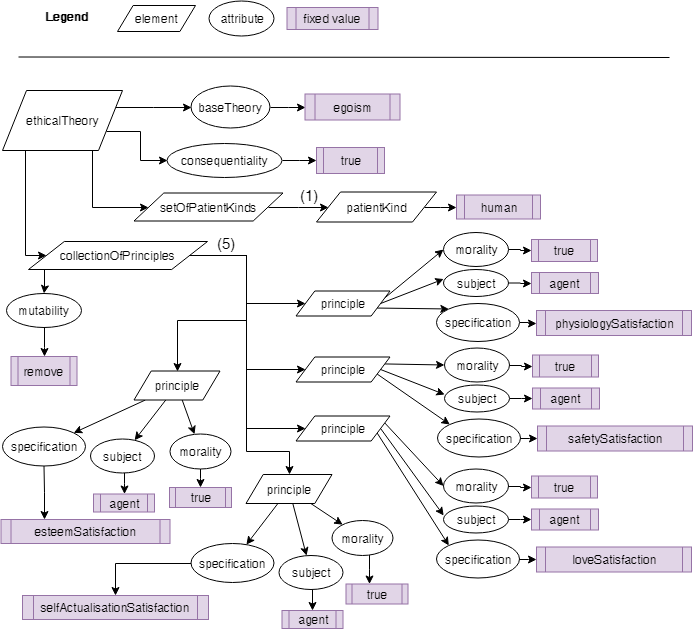}
                \caption{Visualisation of the egoism base model, with unset properties omitted.}
        \label{fig:egobase}
\end{figure}

\begin{figure}[t]
    \centering
    \includegraphics[width=\textwidth]{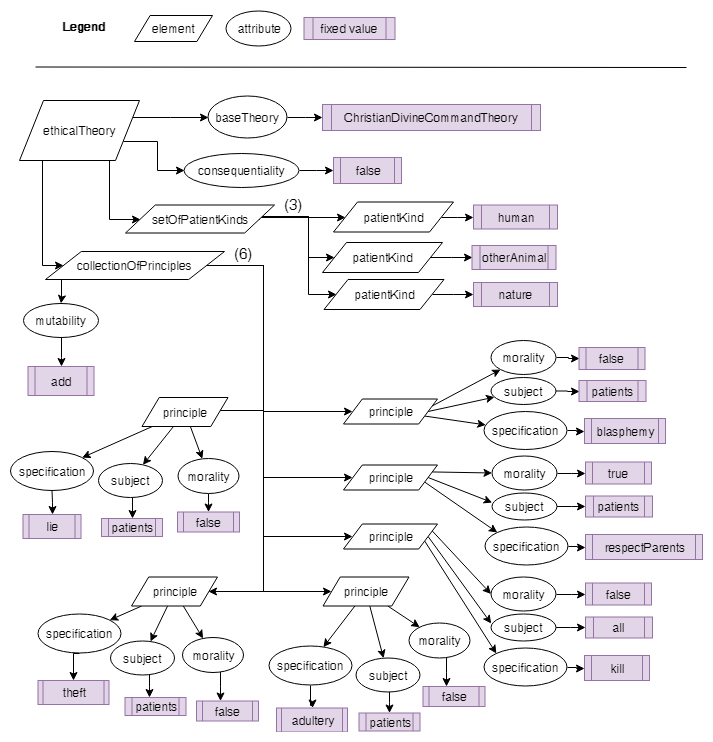}
        \caption{Visualisation of the Christian DCT base model, with unset properties omitted.}
\end{figure}

\clearpage

%\nocite{*}
\bibliographystyle{plain}
\bibliography{DAEcitations}

\end{document}